\documentclass[]{aastex631}
\usepackage{amsmath}
\usepackage{comment} %
\usepackage{subfigure}
\usepackage{booktabs}
\usepackage{multirow} 
\usepackage{makecell}
\usepackage{threeparttable} 
\usepackage{CJK}
\usepackage{makecell}
\usepackage{threeparttable}

\received{----}
\revised{----}
\accepted{17 Dec, 2024}

\submitjournal{ApJ}

\begin{document}
\begin{CJK*}{UTF8}{gbsn}

\title{A New Approach to Identifying Red Supergiant Stars in Metal-poor Galaxies: A Case Study of NGC 6822}

\author[0009-0003-3321-6393]{Zhi-wen Li (李志文)}
\affiliation{Institute for Frontiers in Astronomy and Astrophysics,
            Beijing Normal University,  Beijing 102206, China}
\affiliation{School of Physics and Astronomy,
               Beijing Normal University,
               Beijing 100875, China}

\author[0000-0001-8247-4936]{Ming Yang (杨明)}
\affiliation{Key Laboratory of Space Astronomy and Technology,
             National Astronomical Observatories,
             Chinese Academy of Sciences,
             Beijing 100101, China}

\author[0000-0003-3168-2617]{Biwei Jiang (姜碧沩)}
\affiliation{Institute for Frontiers in Astronomy and Astrophysics,
            Beijing Normal University,  Beijing 102206, China}
\affiliation{School of Physics and Astronomy,
               Beijing Normal University,
               Beijing 100875, China}

\author[0000-0003-1218-8699]{Yi Ren (任逸)}
\affiliation{Department of Astronomy,
            College of Physics and Electronic Engineering,
            Qilu Normal University,
            Jinan 250200, China}

\correspondingauthor{Ming Yang, Biwei Jiang}
\email{myang@nao.cas.cn, bjiang@bnu.edu.cn}



\begin{abstract}
A complete sample of red supergiant stars (RSGs) is important for studying their properties. Identifying RSGs in extragalatic field first requires removing the Galactic foreground dwarfs. The color-color diagram (CCD) method, specifically using $r-z/z-H$ and $J-H/H-K$, has proven successful in several studies. However, in metal-poor galaxies, faint RSGs will mix into the dwarf branch in the CCD and would be removed, leading to an incomplete RSG sample. This work attempts to improve the CCD method in combination with the Gaia astrometric measurement to remove foreground contamination in order to construct a complete RSG sample in metal-poor galaxies. The empirical regions of RSGs in both CCDs are defined and modified by fitting the locations of RSGs in galaxies with a range of metallicity. The metal-poor galaxy NGC 6822 is taken as a case study for its low metallicity ([Fe/H] $\approx$ -1.0) and moderate distance (about 500 kpc). In the complete sample, we identify 1,184 RSG, 1,559 oxygen-rich AGB (O-AGBs), 1,075 carbon-rich AGB (C-AGBs), and 140 extreme AGB (x-AGBs) candidates, with a contamination rate of approximately 20.5\%, 9.7\%, 6.8\%, and 5.0\%, respectively. We also present a pure sample, containing only the sources away from the dwarf branch, which includes 843 RSG, 1,519 O-AGB, 1,059 C-AGB, and 140 x-AGB candidates, with a contamination rate of approximately 6.5\%, 8.8\%, 6.1\%, and 5.0\%, respectively. About 600 and 450 RSG candidates are newly identified in the complete and pure sample, respectively, compared to the previous RSG sample in NGC 6822. 

\end{abstract}



\section{Introduction}

Red supergiant stars (RSGs) are massive, He-burning, population I stars. Generally speaking, the initial masses of RSGs are $\sim 8-30 M_\odot$. They have young ages of $\sim 8-20$ Myr, low effective temperatures of $\sim 3,500-4,500$ K, large radii of $\sim 100-1,000  R_\odot$, and high luminosities of $\sim 4,000-400,000 L_\odot$ \citep{1979ApJ...232..409H,2005ApJ...628..973L,2013EAS....60...31E, 2013NewAR..57...14M,2017ApJ...847..112D}. Massive stars leave the main sequence and then streak horizontally from the upper left to the upper right, evolving into RSGs stage, which shows a nearly vertical distribution on the upper right of color-magnitude diagram (CMD). The faint end of RSGs branch is usually denoted by the tip of red giant branch (TRGB) \citep{2019A&A...629A..91Y,2021A&A...646A.141Y, 2021ApJ...907...18R, 2021ApJ...923..232R, 2020ApJ...900..118N, 2020ApJ...889...44N, 2021AJ....161...79M}.

RSGs are important for chemical enrichment and dust production in interstellar medium due to their high mass loss rate (MLR) and the final-end as supernovae. Although asymptotic giant branch stars (AGBs) contribute the majority of dust in universe, RSGs may dominate the contribution of dust in primitive universe because of their short evolutionary time scale \citep{2005ApJ...634.1286M, 2010NewAR..54....1L}. RSGs in metal-poor galaxies can serve as the probe of primitive low metallicity universe \citep{2000A&ARv..10....1K,2012AJ....144....4M}. One of the final-ends of RSGs is to explode as Type II-P supernovae, which also produces massive amount of dust \citep{2023MNRAS.523.6048S}. This is one of the most drastic and direct actions that affects the surrounding environment and contributes to the birth of next generation stars. 
In fact, there are discrepancies between the derived MLR from different works \citep{2001ApJ...551.1073H, 2010ApJ...717L..62Y, 2011A&A...526A.156M, 2016MNRAS.463.1269B, 2023A&A...676A..84Y}.
Futhermore, RSGs are key to studying the scaling relation between granulation and stellar parameters \citep{2020ApJ...898...24R}, and calibrating the period-luminosity relation \citep{2006MNRAS.372.1721K, 2011ApJ...727...53Y, 2012ApJ...754...35Y, 2018ApJ...859...73S, 2019MNRAS.487.4832C, 2019ApJS..241...35R, 2024IAUS..376..292J}. 

A complete and pure sample of RSGs, that covers a wide range of metallicity, is necessary to answer these challenging questions.  Identifying a complete sample of RSGs within the Galaxy is difficult because of the obstruction of large amount of dust. Meanwhile, the nearby galaxies located away from the Galactic plane and viewed face-on provides the possibility to tackle this problem. The main obstacle of probing such extra-galactic RSGs population is the serious foreground contamination \citep{2007AJ....133.2393M}. The Galactic red dwarfs are blended with RSGs in apparent CMD. Nevertheless, the number of RSGs is increased dramatically, reaching up to over ten thousands in recent years thanks to the advance of observational facilities and invention of new methods \citep{2019A&A...629A..91Y, 2021A&A...646A.141Y, 2021ApJ...907...18R,2021ApJ...923..232R, 2020ApJ...900..118N, 2020ApJ...889...44N, 2021AJ....161...79M}.

There are several methods to distinguish extragalactic RSGs from foreground dwarf stars. \citet{2016ApJ...826..224M} identified 255 RSGs in M31 by radial velocities and spectral features. But spectrum is only available for bright RSGs so that the distance is limited or the faint RSGs may be missed. Besides, the cool star populations like red giants and AGBs share very similar spectral features of RSGs \citep{2024ApJ...965..106Y}, which makes it difficult to separate them spectroscopically.
The color-color diagrams (CCD), such as $B-V/V-R$, $r-z/z-H$, and $J-H/H-K$, provide a valid and efficient way to remove foreground dwarfs. This method makes use of the bifurcation structure in the CCD between low surface gravity ($\log g$) target, i.e., RSGs and AGBs and high $\log g$ targets, i.e., dwarfs.
\citet{1998ApJ...501..153M} successfully distinguished RSGs from foreground dwarfs using the $B-V/V-R$ diagram by the dependence of metallic lines in stellar atmosphere on $\log g$. In accordance with the development in the photometric system of large-scale surveys, new CCDs involving infrared bands are created. Specifically, the $r-z/z-H$ and $J-H/H-K$ diagrams are developed in virtue of more emission in $H$-band (H-bump) by low $\log g$ targets than high $\log g$ targets, which results in the bifurcation in such CCD diagrams.
\citet{2021A&A...647A.167Y} successfully distinguished RSGs from dwarfs and identified 323 RSGs in NGC 6822 by the $r-z/z-H$ diagram.
\citet{2021ApJ...907...18R} removed foreground dwarfs in front of M31 and M33 and identified 5,498 and 3,055 RSGs, respectively, mainly using $J-H/H-K$ diagrams. This significant increase in the number of RSGs demonstrates the effectiveness of the new CCD. Moreover, extinction and varibility are also much smaller in the near-infrared (NIR) bands compared to the optical bands, which is very helpful.
Base on these two new CCD diagrams, \citet{2021ApJ...923..232R} continue to identify a total of about 9,000 RSGs in twelve dwarfs galaxies in the Local Group.

In addition to spectroscopic and photometric measurements, astrometric information such as parallax and proper motion measured by Gaia \citep{2023A&A...674A...1G} are completely new and independent of metallicity, which can be used to effectively remove the foreground stars. \citet{2019A&A...629A..91Y} identified the member stars of LMC and SMC with astrometry as well as radial velocity in the Gaia/DR2 data. On this bases, \citet{2021ApJ...923..232R} later successfully removed foreground contaminants and identified 2,138 and 4,823 RSGs in SMC and LMC respectively with the Gaia/DR3 data and the help of the CCD method.

The CCD diagrams and Gaia astrometric measurement have been proven effective, and more than 10,000 RSGs in the Local Group galaxies are identified. However, both methods have inherent limitations. The faint end of RSGs in metal-poor galaxies will mix into the dwarf branch in the CCD, which makes the sample incomplete. Regarding the Gaia astrometric method, it is very challenging to apply it to galaxies beyond the Magellanic Clouds due to limitations in the observational sensitivity. In order to construct a complete sample of RSGs in metal-poor and distant galaxies, this work attempts to find a new approach by combining CCD method with Gaia astrometry.
As a case study, NGC 6822 is selected for its poor metallicity and moderate distance (m-M = 23.40, 479kpc; \citealt{2012MNRAS.421.2998F}). In this work, we adopt [Fe/H] = -1.0 \citep{2012AJ....144....4M} for NGC 6822, which agrees with the result of -1.0 $\pm$ 0.3 \citep{2003PASP..115..635D}, -1.05 \citep{2022Univ....8..465R}, -1.14 $\pm$ 0.08 \citep{2011ASPC..445..409S}, and $\sim$ -1.286 $\pm$ 0.095 \citep{2020ApJ...892...91H}.
The paper is organized as follows, Sect.\ref{Sect.data_reduction} for the data reduction, Sect.\ref{Sect.method} for the method to remove foreground stars, and Sect.\ref{Sect.identification} for identifying RSGs and AGBs in the CMD. Discussions are given in Sect.\ref{Sect.discussions}, and Sect.\ref{Sect.conclusions} gives the summary.

\section{Data Reduction} \label{Sect.data_reduction}

\begin{figure}
    \centerline{\includegraphics[width=0.45\linewidth]{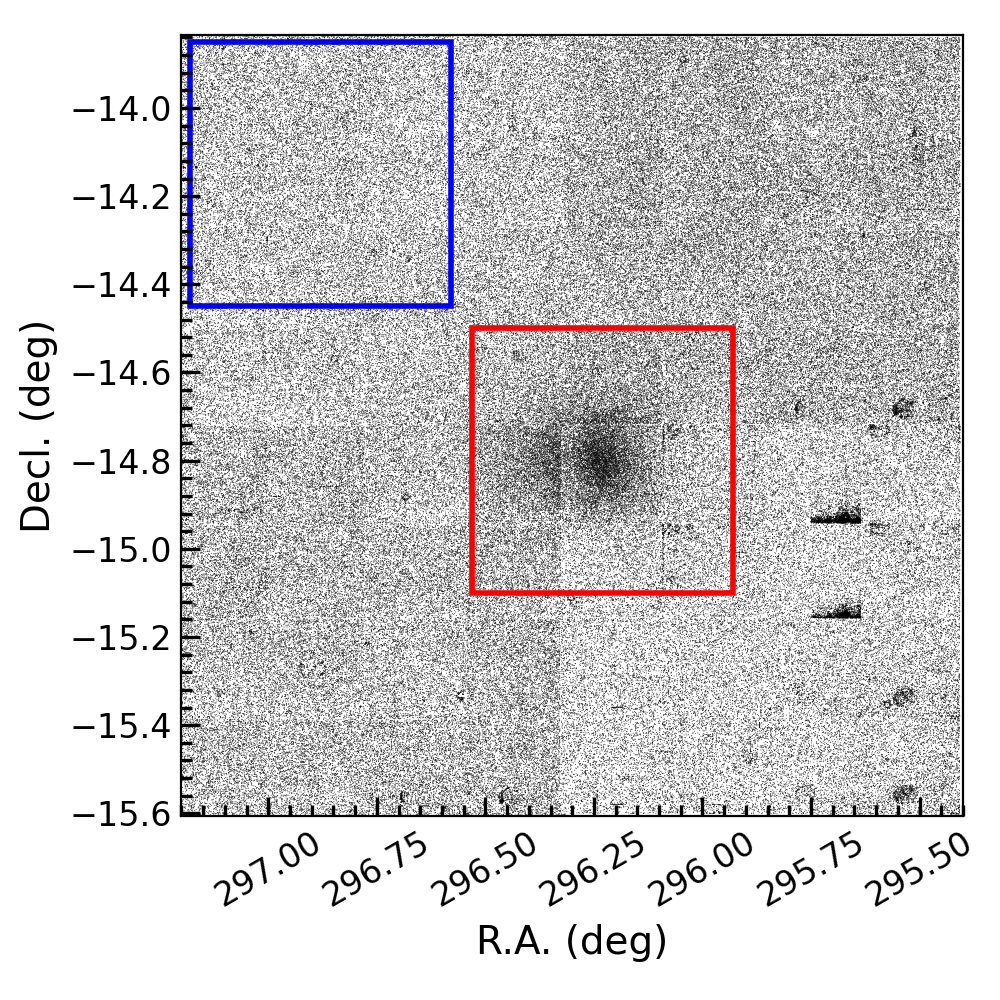}}
    \caption{Field of view of NGC 6822. The black dots denote point sources from UKIRT. The red and blue square represents the sample and reference region, respectively.}
    \label{Figure.field_of_view}
\end{figure}

To combine the CCD and Gaia astrometric method, we collect the optical data in $r$- and $z$-band, NIR data in $J$-, $H$- and $K$-band and astrometric data for NGC 6822. The sky area studied in this work (sample region) is defined by a square with half-side length that is three times of the half-light radius of the galaxy ($r_h = 2.65'$, \citealt{2012AJ....144....4M}), i.e., $295.93^\circ <\text{R.A.}< 296.53^\circ$, $-15.10^\circ <\text{Decl.}<-14.50^\circ$, as shown by the red square in Figure \ref{Figure.field_of_view}.

\subsection{The Pan-STARRS data}
The optical photometric data in $r$- and $z$-band are collected from the Panoramic Survey Telescope and Rapid Response System (Pan-STARRS, PS1; \citealt{2016arXiv161205560C}) DR2, with the limited magnitude of 23.2 mag and 22.3 mag, respectively. Data quality is controlled by: 1) nDetections $\geq$ 2, to remove sources with detections less than twice in a single epoch in all filters, 2) qualityFlag $\neq$ 1, 2 and $<$ 64, to remove extended sources in PS1 (qualityFlag = 1) or in external data (qualityFlag = 2), and low-quality data (qualityFlag $\geq$ 64), and 3) rMeanPSFMag-rMeanKronMag $<$ 0.05, to remove galaxies. There are 40,713 sources selected and  processed further by: 1) self-crossmatching with a search radius of 1$\arcsec$ to remove 14 duplicate sources, and 2) removing the top 3$\%$ sources ranked by photometry error in any bands of $griz$, i.e. 3,834 sources with the largest photometric error. The 36,865 sources are kept.

\subsection{The Near-infrared Data}
The NIR data in $J$-, $H$- and $K$-band come from images taken with the Wide Field Camera (WFCAM) on the 3.8 m United Kingdom Infra-Red Telescope (UKIRT) located in Hawaii from mid-2005 to 2008 \citep{2013ASSP...37..229I}. The details of observation and data processing can be found in \citet{2021ApJ...923..232R}. In brief, the images are processed by the Cambridge Astronomical Survey Unit (CASU) and make available via the WFCAM Science Archive\footnote{\url{http://wsa.roe.ac.uk/}}. The catalogue contains processed photometric data in $J$-, $H$- and $K$-band, applying all necessary corrections provided by CASU\footnote{\url{http://casu.ast.cam.ac.uk/surveys-projects/software-release/fitsio_cat_list.f/view}}. This yields 45,228 sources which are further purified by: 1) self-crossmatching with a search radius of 1$\arcsec$ to remove 24 duplicate sources, 2) removing 15,523 sources lacking photometric data in any of the $J$-, $H$- or $K$-band, and 3) removing 109 sources with photometric SNR $<$ 4 in any bands. The 29,572 sources are kept.

\subsection{Gaia/DR3}
The parallax, proper motion (PM), and photometric data in the $G$-, $BP$- and $RP$-band are collected from Gaia/DR3 \citep{2023A&A...674A...1G}. There are 21,794 sources collected which are processed further to control the data quality by: 1) self-crossmatching with a search radius of 1$\arcsec$ to remove 184 duplicate sources, 2) removing 139 sources with $RUWE$ $>$ 2.0 (Renormalised Unit Weight Error, that equals to 1 ideally), and 3) removing 11 sources with $P_{\rm gal} > 0.5$ (probability as a galaxy). The 21,460 sources are kept.

\section{Removing Foreground Dwarfs}\label{Sect.method}
\begin{figure}
    \centerline{\includegraphics[width=1\linewidth]{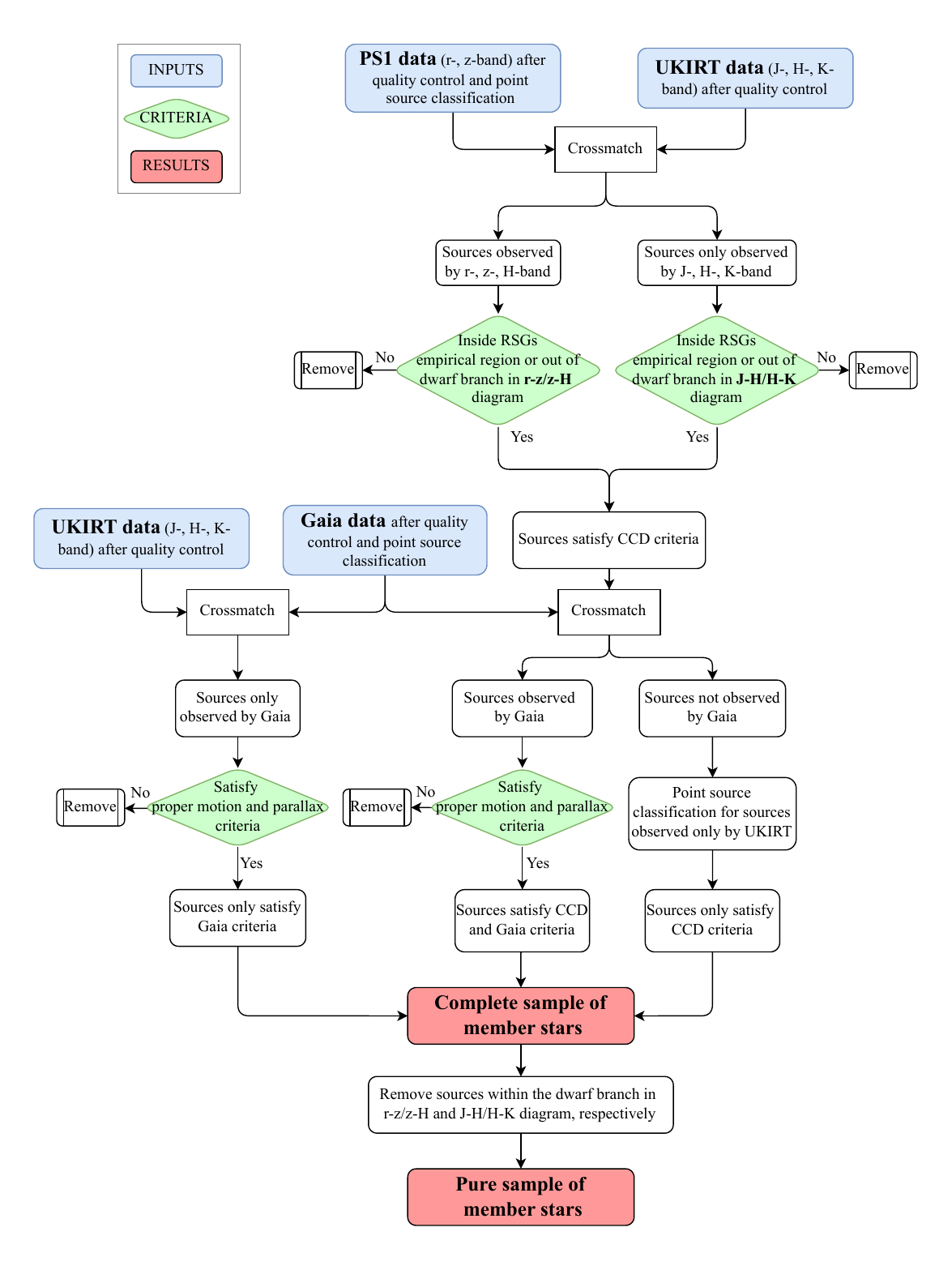}}
    \caption{The flowchart of the whole process for removing foreground dwarfs.}
    \label{Figure.flowchart}
\end{figure}

The foreground dwarfs are first removed with the CCD. The $r-z/z-H$ diagram takes the priority over the $J-H/H-K$ diagram because the bifurcation between low and high $\log g$ targets is more evident in this CCD, which is evident in the work of \citet{2021ApJ...923..232R}. A flow chart of the whole process to remove foreground dwarfs is presented in Figure \ref{Figure.flowchart}, and the details are presented in the following text.

\subsection{The $(r-z)_{0}/(z-H)_{0}$ Diagram}\label{Sect.rzh}
\subsubsection{The Dwarf Branch}
\begin{figure}
    \centerline{\includegraphics[width=0.6\linewidth]{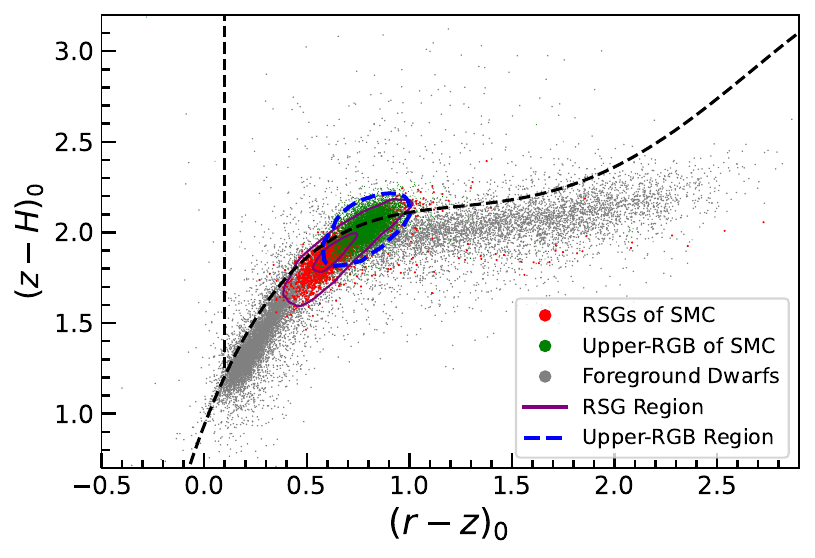}}
    \caption{The $(r-z)_0)/(z-H)_0$ diagram from the reference region of NGC 6822. The gray dots represent the foreground Galactic dwarfs within the reference region. The red and green dots represent the RSGs and the brightest 0.7 magnitude RGBs in SMC, with the purple and blue dashed circle represent their boundaries, respectively. The curved black dashed line represents the borderline of the dwarf branch, and the vertical dashed line represents the blue border of the RSGs in $(r-z)_0$. }
    \label{Figure.ccd_rzh}
\end{figure}

As mentioned earlier, the faint end of RSGs in a metal-poor galaxy is mixed with the Galactic dwarf stars in the CCD. Thus, an independent reference area (reference region) with a size similar to NGC 6822 as shown by the blue square in Figure \ref{Figure.field_of_view} is selected to define the dwarf branch in the CCD. This area will also be used to calculate the contamination rate of the RSG sample later. The sources in the $(r - z)_0/(z - H)_0$ diagram are displayed as gray dots in Figure \ref{Figure.ccd_rzh}, where the intrinsic color index is calculated by subtracting a uniform foreground extinction of $E(B-V)$ = 0.169 mag based on the extinction map of \citet{1998ApJ...500..525S} with the conversion coefficients under $R_V$ = 3.1 from \citet{2011ApJ...737..103S}. The borderline of the Galactic dwarf branch is obtained by calculating the ridge and the width of the distribution, which is shown by the black dashed line in Figure \ref{Figure.ccd_rzh}.

\subsubsection{The RSG region}
Different from the dwarf stars, the location of metal-poor RSGs in the CCD is not so well studied. We try to find out their locations with stellar atmospheric models. The calculation is performed by convolving spectral energy distribution of typical metal-poor RSGs from both the MARCS \citep{2008A&A...486..951G} and ATLAS9 \citep{2003IAUS..210P.A20C} code with the corresponding filter transmission curves. 
For the SMC metallicity, the results from the models deviate apparently from the locations of the observational RSG sample of SMC in \citet{2021ApJ...923..232R}. In this work, we adopt the [Fe/H] = -1.0 \citep{2012AJ....144....4M} for SMC, which agrees with the result of -0.95 $\pm$ 0.08 \citep{2018MNRAS.475.4279C}, -0.99 $\pm$ 0.01 \citep{2014MNRAS.442.1680D}, and $\sim$ -1.2 \citep{2024AJ....167..123L}. Furthermore, the change of the model RSG locations with metallicity in the CCD is inconsistent with the observation of LMC, M33, and M31.
This discrepancy can be understood by the uncertainty in the RSG atmospheric models. Thus, the location of RSGs in the CCD has to be determined empirically, for which the RSG regions at different metallicities are entirely derived based on the RSGs in the SMC, and subsequently shifted and rotated to fit RSG populations in other galaxies (e.g., the LMC, M31, and M33). The process is in three steps as described below.

First, an empirical region of RSGs in the $(r-z)_0/(z-H)_0$ diagram is determined by contouring the RSGs in the SMC that are selected from \citet{2021ApJ...923..232R}, because the SMC is the most metal-poor one among the four galaxies (SMC, LMC, M33, and M31) with the most complete RSG sample. Due to that the PS1 survey does not cover the SMC area, the $r$- and $z$-band data are retrieved by crossmatching with a search radius of 1$''$ with Sky Mapper DR2 \citep{2019PASA...36...33O}. The foreground extinction is corrected by using the extinction map of \citet{2021ApJS..252...23S}. We adopted the extinction coefficients from \citet{2019ApJ...877..116W} with $R_V$ = 3.16, considering the extinction map includes the foreground extinction of the Milky Way. Moreover, due to the small mean extinction values of Magellanic Clouds ($E(V-I) = 0.047$ for the SMC and 0.100 for the LMC) as reported by \citet{2021ApJS..252...23S}, the impact of using different $R_V$ values is negligible. As shown in Figure \ref{Figure.ccd_rzh}, the outermost purple solid line represents the empirical region of RSGs, where the marginal density of the contour is 5\% of the maximum. In practice, this region is appropriately enlarged by a factor of 1.3 to compensate for the extinction and photometric error in the optical bands. In addition, to define the TRGB which denotes the faint end of RSGs, the region of the upper-RGB in the CCD is determined by contouring the brightest 0.7 magnitude RGBs in the SMC, as shown by the blue dashed circle in Figure \ref{Figure.ccd_rzh}.

Second, the effect of metallicity on the RSG region in the CCD is investigated. Previous studies have obtained more or less complete sample of RSGs in the SMC, LMC, M33 and M31, which span a wide metallcity range from about -1.0 to +0.3. The adopted [Fe/H] for LMC is -0.5 \citep{2012AJ....144....4M}. According to \citet{1995RMxAC...3..133E} and \citet{1997ApJ...489...63G}, M33 has a comparable value of 12+log[O/H] to the Milky Way in that 8.70 for the Milky Way and 8.75 for M33. Therefore, we adopt [Fe/H] = 0.0 for M33. In addition, [Fe/H] = 0.3 is taken for M31 from  \citet{2018MNRAS.478.5379D}.

\begin{figure}
    \centerline{\includegraphics[width=0.8\linewidth]{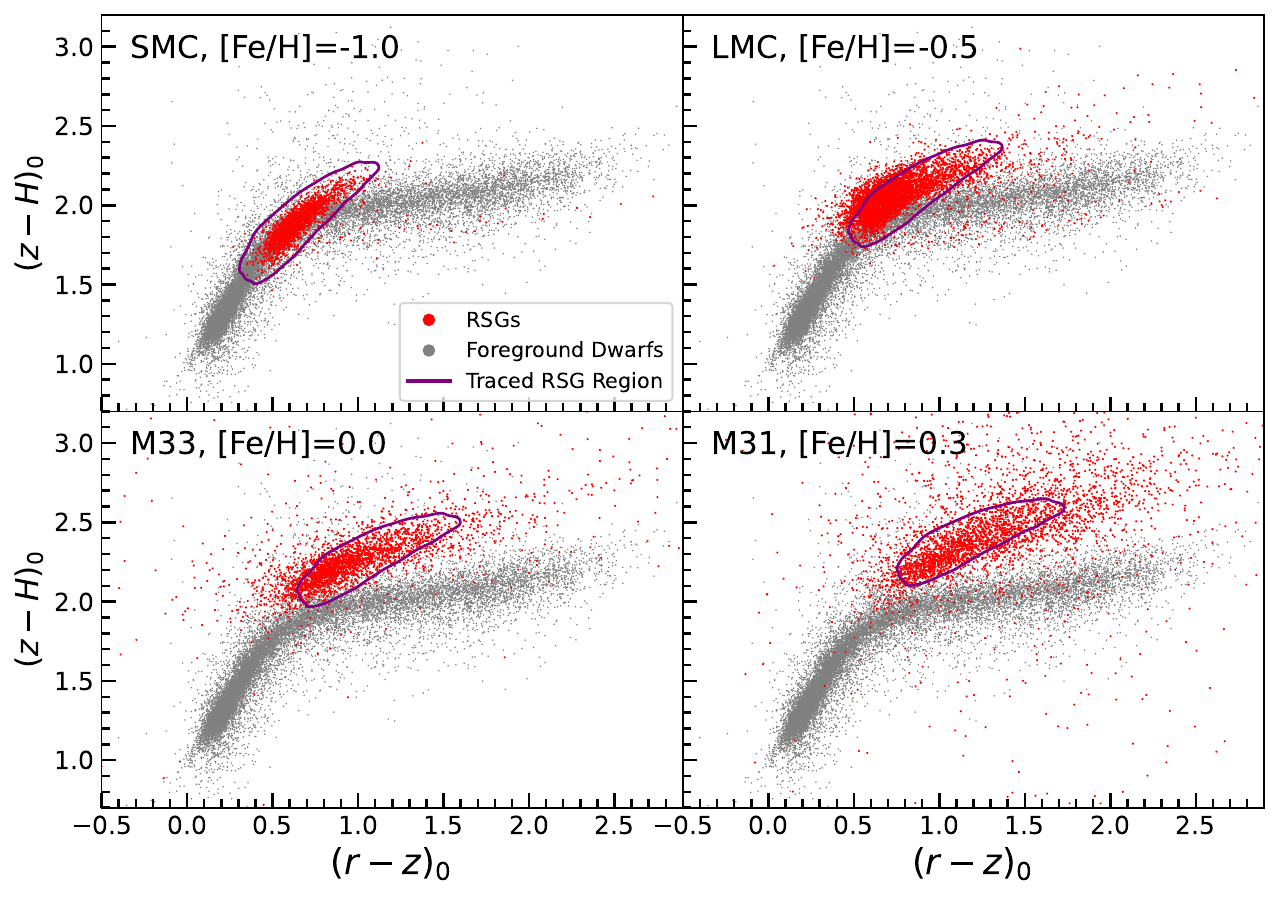}}
    \caption{The variation of the RSG region with [Fe/H] in the $(r-z)_0)/(z-H)_0$ diagram. The red dots represent RSGs in the SMC, LMC, M31, and M33 \citep{2021ApJ...907...18R,2021ApJ...923..232R}, respectively, and the gray dots represent foreground Galactic dwarfs. The purple circle marks the empirical region of RSGs, derived by contouring the RSGs in the SMC (where the marginal density of the contour is 5\% of the maximum) and subsequently shifting and rotating it using the metallicity relations in Eq.\eqref{eq.optical_color_feh} and Eq.\eqref{eq.theta_feh}.}
    \label{Figure.four_galaxies}
\end{figure}

There are 2,138, 4,823, 3,055, and 5,498 RSGs in SMC, LMC, M33, and M31, respectively, in \citet{2021ApJ...907...18R,2021ApJ...923..232R}. It should be mentioned that the $r$- and $z$-band data of LMC are also retrieved from Sky Mapper DR2 \citep{2019PASA...36...33O}. For LMC, the same extinction map and law are adopted as for the SMC, as mentioned above. For M33 and M31, the extinction is uniformly corrected with the mean $A_V\sim$ 0.11 mag and 0.17 mag \citep{1998ApJ...500..525S}, respectively. 
The red dots in Figure \ref{Figure.four_galaxies} show the locations of RSGs in the $(r-z)_0/(z-H)_0$ diagram for the four galaxies.

\begin{figure}
    \centerline{\includegraphics[width=0.8\linewidth]{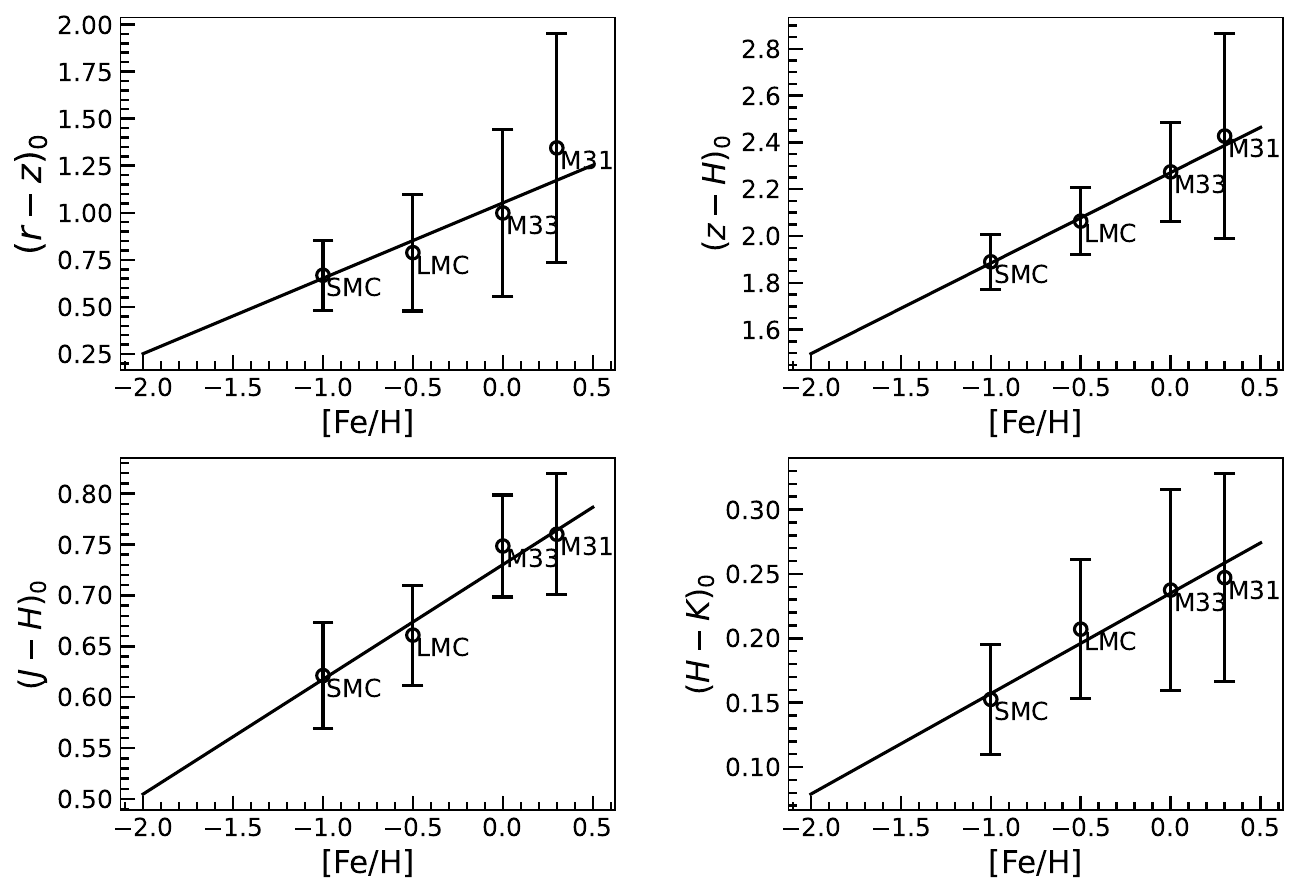}}
    \caption{The variation of the intrinsic colors of RSGs  with [Fe/H]. The error bars represent the 1$\sigma$ dispersion of the color, and the solid line is the result of a linear fitting.}
    \label{Figure.color_feh}
\end{figure}

The relation of $(r-z)_0$ and $(z-H)_0$ of RSGs with [Fe/H] is then fitted with these observational data. The mean color and its dispersion are calculated for all RSGs in each galaxy, and a linear function is used to fit the variation of the color index with metallicity. As shown in Figure \ref{Figure.color_feh}, the color of RSGs becomes bluer as [Fe/H] decreases. The fitting results are as following:
\begin{equation}
\begin{aligned}
    (r-z)_0 &= 0.400[Fe/H] + 1.053 \\
    (z-H)_0 &= 0.387[Fe/H] + 2.271
\end{aligned}
\label{eq.optical_color_feh}
\end{equation}

\begin{figure}
    \centerline{\includegraphics[width=0.5\linewidth]{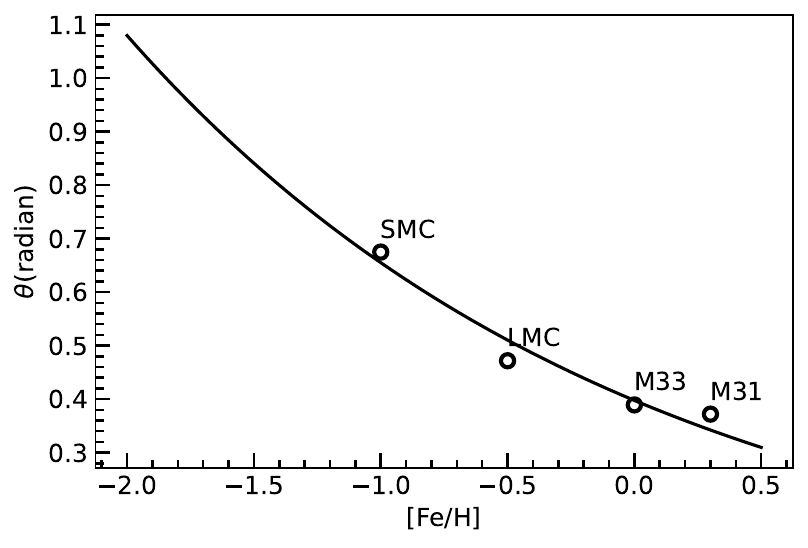}}
    \caption{An exponential fitting of the inclination angle of the RSG branch in the $(r-z)_0/(z-H)_0$ diagram with [Fe/H].}
    \label{Figure.rotate_feh}
\end{figure}

Third, a rotation parameter is considered when shifting the empirical region of RSGs along with [Fe/H]. It is noted that there is a change in the inclination angle of the RSGs branch with metallicity in the CCD, such that the RSGs branch becomes more vertical as metallicity decreases, visible in Figure \ref{Figure.four_galaxies}. The inclination angle ($\theta$) of the RSGs branch in the CCD for each galaxy is determined by fitting the distribution of RSGs with a linear function, where $k=\tan\theta$. The values of $k$ for the SMC, LMC, M33, and M31 are 0.80, 0.51, 0.41, and 0.39, respectively, corresponding to $\theta$ = 38$\arcdeg$, 27$\arcdeg$, 22$\arcdeg$, and 21$\arcdeg$. Then the relation between $\theta$ and [Fe/H] is fitted by an exponential function, which yields
\begin{equation}
    \theta = 0.397 \exp(-0.500[Fe/H])
\label{eq.theta_feh}
\end{equation}
The result is illustrated in Figure \ref{Figure.rotate_feh}.

According to Eq.\eqref{eq.optical_color_feh} and Eq.\eqref{eq.theta_feh}, the required shift and rotation of the empirical region of RSGs can be calculated from the difference of metallicity between the target galaxy and the SMC. As shown in Figure \ref{Figure.four_galaxies}, the traced empirical region of RSGs (purple circle) fits well with the observational RSGs in different galaxies in the CCD. The upper-RGB region is also shifted with Eq.\eqref{eq.optical_color_feh}, but with no rotation as there is no distinct variation with metallicity of the RGB branch in the CCD. 

We note that this color$/\theta$ - [Fe/H] relation is derived using the average [Fe/H] of the galaxies rather than the [Fe/H] of RSGs. Typically, the latter one is higher than the former one since RSGs are younger stars. For example, \citet{2015ApJ...803...14P} reported a mean [Fe/H] of -0.52 $\pm$ 0.21 for 11 bright RSGs in NGC 6822, which is higher than the value of -1.0 adopted in this work. However, studies on RSG metallicities in extragalactic systems are relatively limited, and obtaining sufficient data from the existing literature is challenging. Therefore, we use the more readily available metallicity of the galaxies to indirectly infer the relation between RSGs and metallicity. This approach represents a necessary compromise given the current constraints. For NGC 6822, due to the similarity in metallicity between young stars in NGC 6822 and the SMC \citep{2015ApJ...806...21D}, this difference does not significantly affect our analysis. 

\subsubsection{Removing Foreground Dwarfs in the $(r-z)_{0}/(z-H)_{0}$ diagram}
For NGC 6822, there are 16,421 sources that have all the $r$-, $z$- and $H$-band data by crossmatching PS1 with the UKIRT data with a search radius of 1$\arcsec$. The selection criteria in $(r-z)_{0}/(z-H)_{0}$ diagram are as follows: 1) Sources within the empirical region of RSGs are kept, even if they are mixed into the dwarf branch, in order to ensure the completeness of potential RSGs, 2) Sources within the upper-RGB region are kept, and 3) Sources below the borderline of the dwarf branch or $(r-z)_0\leq0.1$ are removed. 
As a result, 5,394 sources are kept. As shown in Figure \ref{Figure.rzh_selection}, the orange dots represent the sources within the empirical region of RSGs, but overlapped with the dwarf branch, and the red dots indicate sources that are not contaminated by dwarfs.

\subsection{The J-H/H-K Diagram}
Sources with no $r$- or $z$-band measurement but with the $J$-, $H$- and $K$-band data are cleaned in the $(J-H)_0/(H-K)_0$ diagram. The procedure is similar to that of the $(r-z)_0$ vs. $(z-H)_0$ diagram. The location of foreground dwarf stars in this NIR CCD is defined by the reference region, which is delineated by the purple dashed line in Figure \ref{Figure.jhk_selection}. The empirical region of RSGs is also determined by contouring the RSGs in SMC as shown by the purple solid line in Figure \ref{Figure.jhk_selection}, and the color relation with [Fe/H] from SMC, LMC, M33, and M31 is as following:
\begin{equation}
\begin{aligned}
    (J-H)_0 &= 0.113[Fe/H] + 0.730 \\
    (H-K)_0 &= 0.078[Fe/H] + 0.235
\end{aligned}
\label{eq.color_feh}
\end{equation}

Meanwhile, no rotation of the RSG region is applied, as it is not prominent in the NIR CCD. Additionally, the upper-RGB region is not separately considered because it almost overlaps with the RSGs in the NIR CCD.

The criteria of selecting RSGs in the $(J-H)_0/(H-K)_0$ diagram are as follows: 1) Sources within the empirical region of RSGs are kept, and 2) Sources outside the RSG region but within the dwarf branch are removed. In addition, sources with $(J-H)_0\leq0.5$ or $(H-K)_0\leq0.0$ are removed because they are apparently bluer than RSGs.

Among the 7,722 sources that only have $J$-, $H$- and $K$-band data, 4,097 sources are kept after the selection, as shown by the orange and red dots in Figure \ref{Figure.jhk_selection}, and the color convention is the same as in Figure \ref{Figure.rzh_selection}.

\subsection{The Gaia Astrometric Criterion}
\begin{figure}
    \centerline{\includegraphics[width=0.8\linewidth]{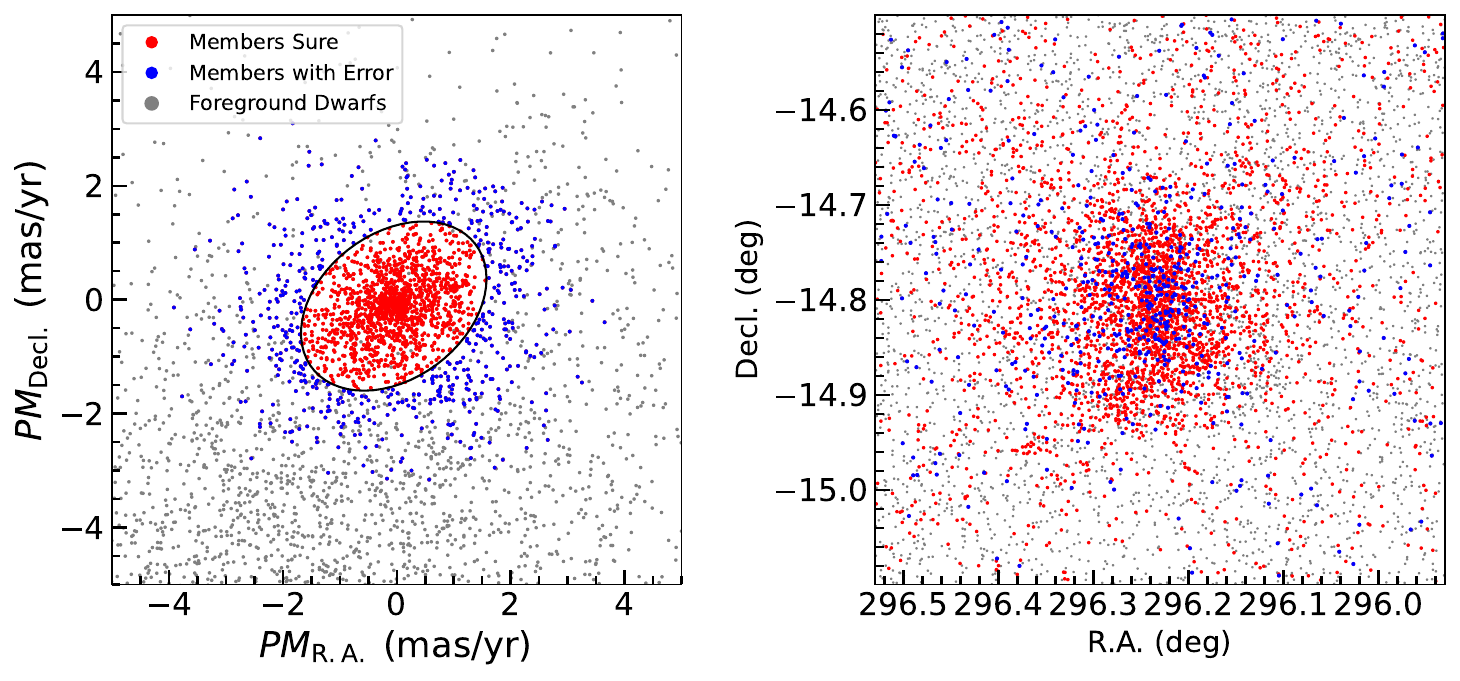}}
    \caption{Left panel: Distribution of the PMs for the Gaia sources that satisfy the CCD criteria. The ellipse represents the constraints of the PM. The red dots represent sources that fall within the PM ellipse, the blue dots represent sources that fall within the PM ellipse after accounting for the error, and the grey dots represent the removed sources. 
    Right panel: Spatial distribution of the sources from the left panel.}
    \label{Figure.PM_selection}
\end{figure}

The fraction of Gaia detection decreases with the distance to the galaxy due to the limiting magnitude in $G$-band being about 20.5 mag. No other galaxies are like the Magellanic Clouds to own an almost complete sample of RSGs by Gaia, still some foreground stars can be removed by the Gaia astrometric information. Besides, crowding is indeed an important factor due to Gaia's small aperture (1.45 m $\times$ 0.5 m). However, for NGC 6822, since RSGs are among the brightest stellar populations and brighter than the limiting magnitude of Gaia (with the faint end brighter than $RP_0$ = 19.5 mag), the impact of crowding is relatively limited. This is further supported by subsequent JWST images presented in the following sections. For NGC 6822, the concentration of member stars are visible in the $PM_\text{R.A.}$ vs. $PM_\text{Decl.}$ diagram of the 21,460 sources detected by Gaia in the sample region as shown in Figure \ref{Figure.PM_selection}. An ellipse is fitted to the distribution, with center at $PM_\text{R.A.}^{\rm center}=-0.05$ mas/yr and $PM_\text{Decl.}^{\rm center}=-0.11$ mas/yr. The semi-major and semi-minor axes are 1.80 mas/yr and 1.26 mas/yr which are three times of the standard deviation, with a position angle of 37$^\circ$. Sources that fall within the PM ellipse are kept. Besides, the sources that fall within the PM ellipse after accounting for 1$\sigma$ measurement error are also kept as potential member stars. The PM selection is shown in Figure \ref{Figure.PM_selection}. Moreover, sources with a distance smaller than 5 kpc calculated by using Gaia parallax and relative error less than 20\% are removed. All the sources observed by Gaia but without astrometric data are kept, because they are likely to be the member stars of the distant target galaxy. 
Among the 14,861 sources that satisfy the CCD criteria, 7,455 are detected by Gaia. After the Gaia astrometric selection, 3,741 of these sources are kept, which constitute the sample of member stars that satisfy both the CCD and Gaia astrometric criteria, denoted as CCD-Gaia sample hereafter.

\subsection{Two Supplementary Samples of Member Stars}
In addition to the CCD-Gaia sample, there are two supplementary ones of member stars. One sample consists of 7,406 sources that satisfy the CCD criteria but have no Gaia data. A further selection of point sources is carried out because no external stellar or extended classification flags from Gaia or PS1 can be used as the other sample. They are required to have the stellar classification flag of -1 (stellar), -2 (probably stellar) or -7 (source with bad pixels) in at least two of the $J$-, $H$-, $K$-bands in the UKIRT data. A total of 6,339 sources are selected, as CCD-only sample.
On the contrary, the other sample consists of 4,287 stars that have the Gaia data but no UKIRT data, for which most of them are blue or faint sources. This Gaia-only sample consists of 2,720 stars after the Gaia astrometric selection.

The combination of the three samples, CCD-Gaia, CCD-only, and Gaia-only consists the complete sample of the member stars. 

\section{Identifying Red Supergiant Stars in the Color-magnitude Diagram}
\label{Sect.identification}
After removing the foreground dwarfs by using the above methods, three independent samples of member stars are collected. The RSGs can then be identified in the CMD from their distinguished color and brightness. It should be noted that AGBs are close to RSGs in the CMD and may contaminate the RSG sample at the faint end because of photometric error and inhomogeneous extinction of the host galaxy. For the stars in the CCD-Gaia and CCD-only sample, the stellar type is identified in the $(J-K)_0$ vs. $K_0$ diagram, while for the stars in the Gaia-only sample, it is identified in the $(BP-RP)_0$ vs. $RP_0$ diagram.

\subsection{The $(J-K)_0$ vs. $K_0$ Diagram}
For either CMD, TRGB is a key point because both RSGs and AGBs are brighter than TRGB as demonstrated in previous works (\citealt{2006A&A...448...77C, 2011ASPC..445..473B, 2019A&A...629A..91Y, 2021ApJ...907...18R, 2021ApJ...923..232R, 2020ApJ...900..118N, 2020ApJ...889...44N, 2021AJ....161...79M}). Here the TRGB is determined in the $(J-K)_0$ vs. $K_0$ diagram by the Sobel-filter edge detection method that only detects the $K$-band magnitude of TRGB \citep{1993ApJ...417..553L,1996ApJ...461..713S, 2018AJ....156..278G, 2018ApJ...858...12H,2023ApJ...954...87W}. This yields the position of TRGB at $K_0 = 17.41$ ($K$-TRGB hereafter) for NGC 6822, which agrees very well with the result of \citet{2011ASPC..445..409S} at $K_0=17.41$ and \citet{2021ApJ...923..232R} at $K_0 = 17.38$.

\begin{figure}
    \centerline{\includegraphics[width=0.79\linewidth]{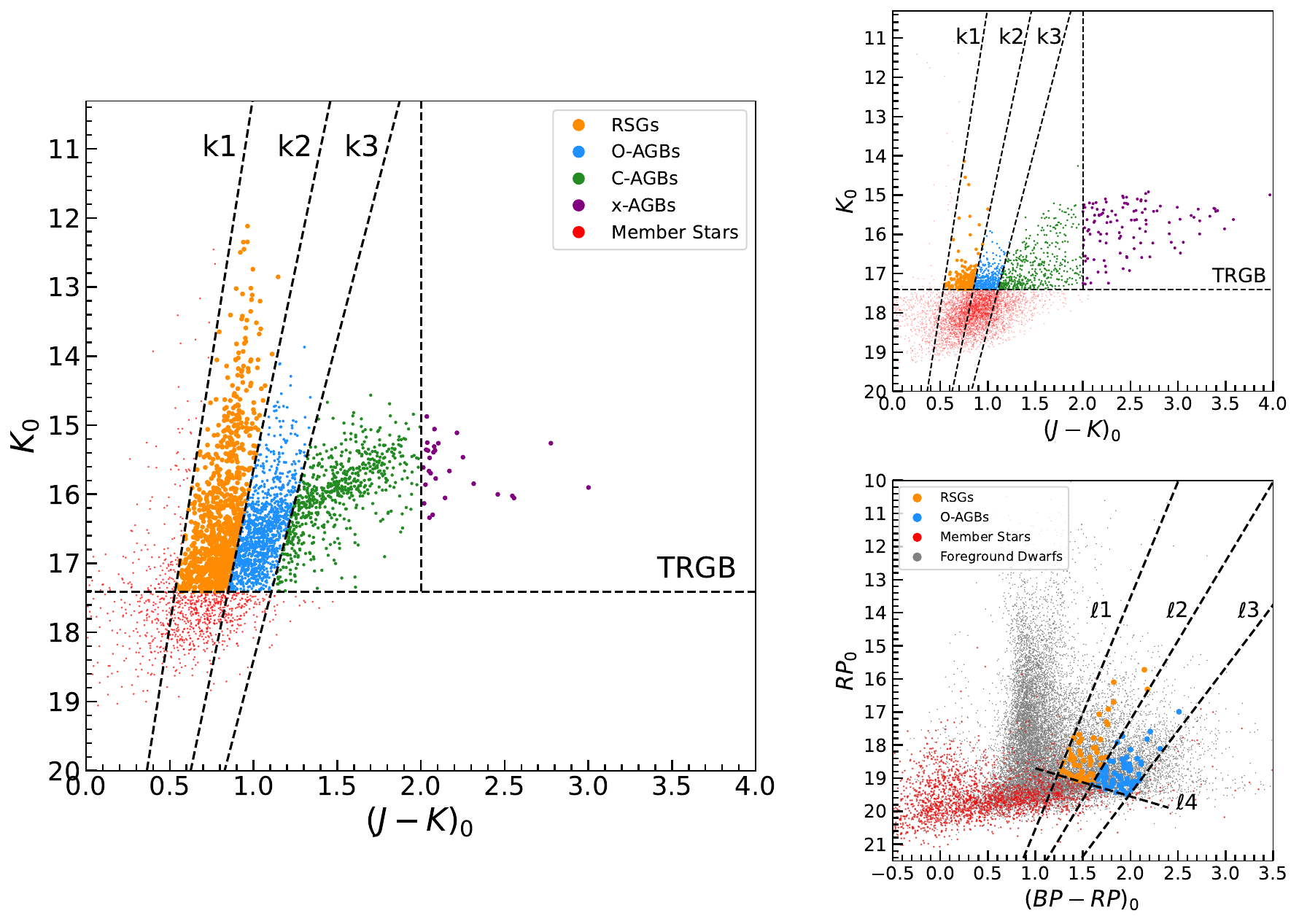}}
    \caption{Identified RSG and AGB candidates of the complete sample in the NIR and optical CMDs.
    The left, upper right, and lower right panels show the sources from the CCD-Gaia, CCD-only, and Gaia-only sample, respectively.
    The orange, blue, green, and purple dots represent RSG, O-AGB, C-AGB, and x-AGB candidates. The red dots represent the rest member stars of NGC 6822, and the gray dots (only in the lower right panel) represent foreground stars.}
    \label{Figure.complete_sample}
\end{figure}

The boundaries of RSGs and AGBs in the $(J-K)_0$ vs. $K_0$ diagram have been determined by several works (\citealt{2006A&A...448...77C, 2011ASPC..445..473B, 2019A&A...629A..91Y}). Here we adopt the recent results from \citet{2022Univ....8..465R}, which are based on the largest sample of RSGs and AGBs in fourteen Local Group galaxies with various metellicity. The three main borderlines of $k1$, $k2$ and $k3$ are manually shifted by eye to match the expected morphological distribution of the stellar populations in the CMD, and specifically listed below (c.f. Figure \ref{Figure.complete_sample}):
\begin{equation}
\begin{aligned}
    k1: K_0 &= -15.366((J-K)_0-0.23) + 22.048 \\
    k2: K_0 &= -11.618((J-K)_0-0.46) + 21.917 \\
    k3: K_0 &= -9.268((J-K)_0-0.56) + 22.481
\end{aligned}
\label{eq.cut_lines_1}
\end{equation}
The RSGs are defined by $k1<K_0\leq k2$ and $K_0<$ $K$-TRGB, O-AGBs by $k2<K_0\leq k3$ and $K_0<$ $K$-TRGB, C-AGBs by $K_0>k3$ and $K_0<$ $K$-TRGB and $(J-K)_0\leq 2.0$, and x-AGBs by $(J-K)_0>2.0$ and $K_0<$ $K$-TRGB.
With the $K_0$ magnitude of TRGB and the above criteria in the $(J-K)_0$ vs. $K_0$ diagram, 959 RSG, 1,154 O-AGB, 620 C-AGB, and 28 x-AGB candidates are identified from the CCD-Gaia sample, and their distribution in the CMD is shown in the left panel of Figure \ref{Figure.complete_sample}. Meanwhile, 163 RSG, 335 O-AGB, 455 C-AGB, and 112 x-AGB candidates are identified from the CCD-only sample whose distribution in the CMD is shown in the upper right panel of Figure \ref{Figure.complete_sample}.

\subsection{The $(BP-RP)_0$ vs. ${RP}_0$ Diagram}
The boundaries of RSGs and AGBs in the ${(BP-RP)}_0$ vs. ${RP}_0$ diagram are defined based on the member stars of SMC \citep{2021ApJ...923..232R} since the metallcity of NGC 6822 is comparable to the SMC's. The boundaries are listed in Eq.\eqref{eq.cut_lines_2}, where $l1$, $l2$ and $l3$ denote the boundary between RSGs and O-AGBs, while $l4$ denotes the TRGB in the $(BP-RP)_0$ vs. $RP_0$ diagram. Besides, $\delta(BP-RP)$ and $\delta RP$ are the color and magnitude shift to account for the difference caused by metallicity, photometry uncertainty, extinction correction, distance, and so on in an individual galaxy, which is 0.18 mag redder and 4.28 mag fainter respectively, to match with the distribution of RSGs and AGBs in NGC 6822.  
\begin{equation}
\begin{aligned}
    l1: RP_0 &= -7.00[(BP-RP)_0+\delta(BP-RP)] + [22.00 +\delta RP] \\
    l2: RP_0 &= -4.80[(BP-RP)_0+\delta(BP-RP)] + [21.70 +\delta RP] \\
    l3: RP_0 &= -3.80[(BP-RP)_0+\delta(BP-RP)] + [22.10 +\delta RP] \\
    l4: RP_0 &= 0.85[(BP-RP)_0-\delta(BP-RP)] + [13.42 +\delta RP]
\end{aligned}
\label{eq.cut_lines_2}
\end{equation}

Our classification criteria in the ${(BP-RP)}_0$ vs. ${RP}_0$ diagram are as follows. RSGs are defined by $l1<RP_0 \leq l2$ and  $RP_0<l4$, and O-AGBs by $l2<RP_0 \leq l3$ and $RP_0<l4$. Only RSGs and O-AGBs are identified in the $(BP-RP)_0$ vs. ${RP}_0$ diagram, as the distribution of other AGBs is less distinct in this CMD. In total, 62 RSG and 70 O-AGB candidates are identified from the Gaia-only sample, and their distribution in the CMD is shown in the lower right panel of Figure \ref{Figure.complete_sample}.

\subsection{Results}
\begin{figure}
    \centerline{\includegraphics[width=1.\linewidth]{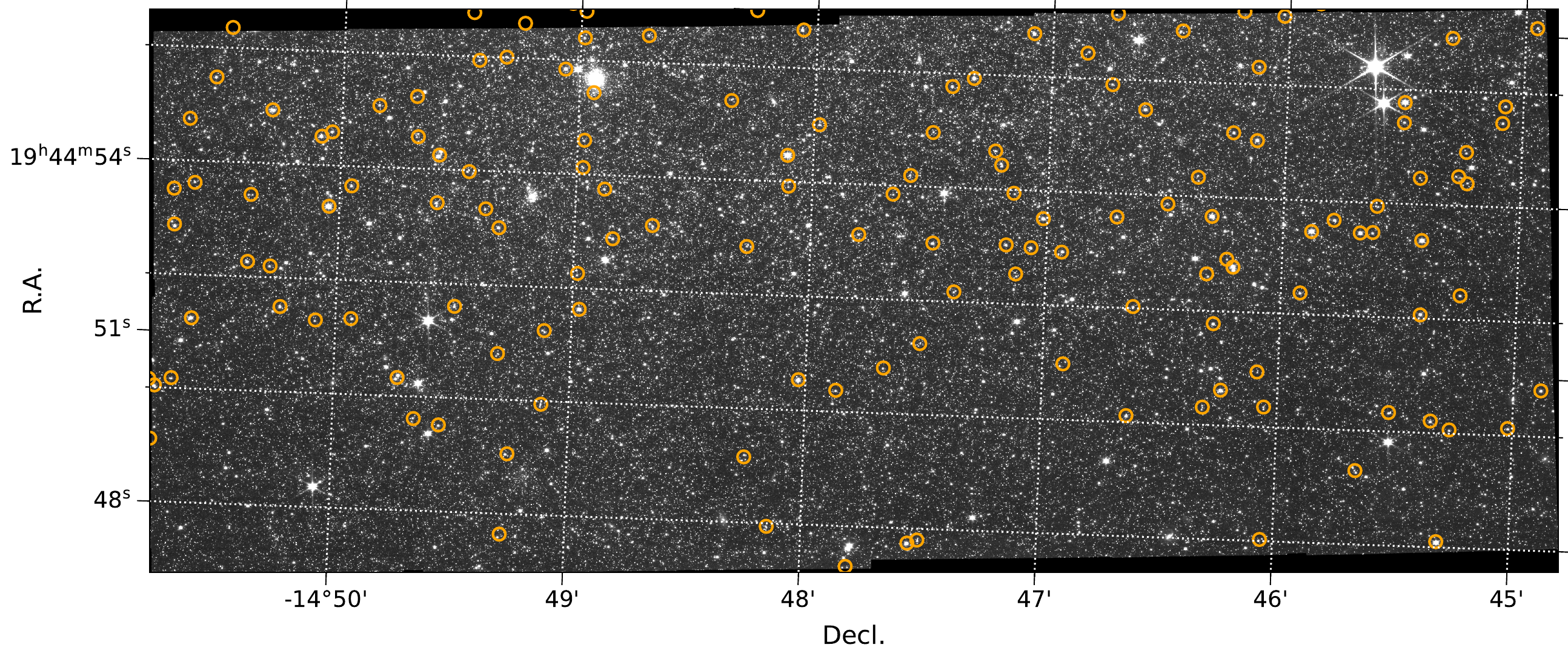}}
    \caption{A JWST image of a patch of sky in NGC 6822. A fraction of RSG candidates in the complete sample is denoted by orange circles.}
    \label{Figure.RSGs_by_JWST}
\end{figure}

As described above, a complete sample of RSGs is constructed by combining all the three sub-samples. At last, there are a total of 1,184 RSG candidates with 959, 163 and 62 from the CCD-Gaia, CCD-only, and Gaia-only sample, respectively. Meanwhile, there are 1,559 O-AGB candidates (1,154, 335, and 70 from the CCD-Gaia, CCD-only and Gaia-only sample, respectively), 1,075 C-AGB candidates (620 and 455 from the CCD-Gaia and CCD-only sample, respectively) and 140 x-AGB candidates (28 and 112 from the CCD-Gaia and CCD-only sample, respectively). The number of identified RSGs and AGBs candidates are listed in Table.\ref{Table.final_sample}. Recently, the James Webb Space Telescope (JWST)\footnote{\url{https://www.stsci.edu/cgi-bin/get-proposal-info?id=1234&observatory=JWST}} captured a patch of the NGC 6822 region in the filter of F115W. With the powerful spatial resolution of JWST, we can clearly see a fraction of RSGs in our complete sample, as shown in Figure \ref{Figure.RSGs_by_JWST} where the identified RSG candidates are denoted by the orange circles. It shows that most of the RSGs are not blended with other objects and should be correctly identified.

\begin{figure}
    \centerline{\includegraphics[width=0.6\linewidth]{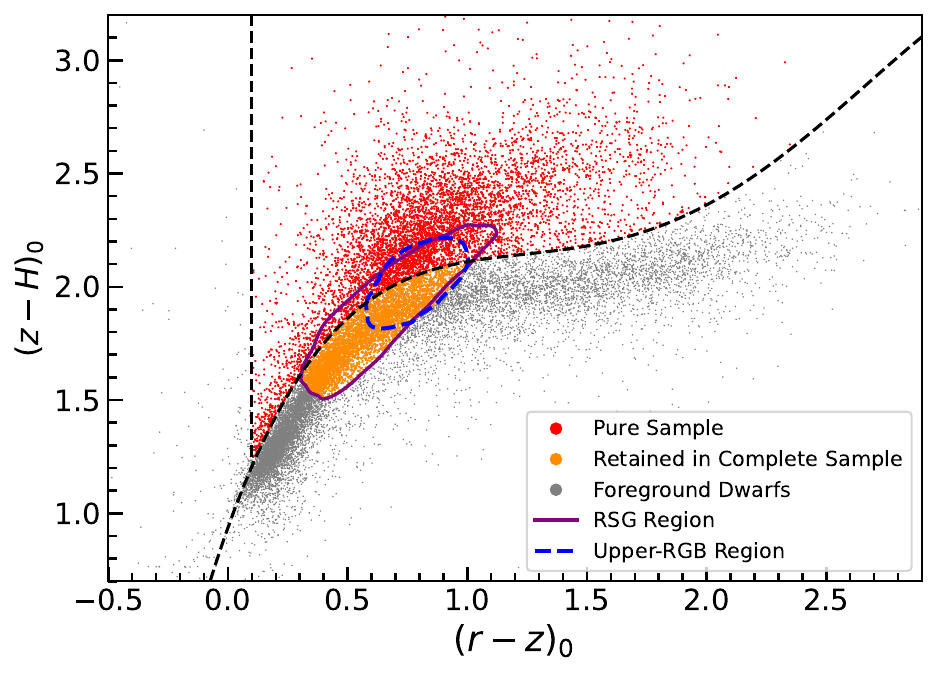}}
    \caption{Selection of RSGs in the $(r-z)_0/(z-H)_0$ diagram. These are for NGC 6822, but with the SMC regions highlighted.
    The purple circle represents the empirical region of RSG, the blue dashed circle represents the upper-RGB region, and the black dashed line denotes the borderline between the foreground dwarfs and the member stars.
    The gray dots represent the foreground dwarfs, the red dots for the certain member stars, and the orange dots for the sources included in the complete sample but not in the pure sample.
    }
    \label{Figure.rzh_selection}
\end{figure}

\begin{figure}
    \centerline{\includegraphics[width=0.6\linewidth]{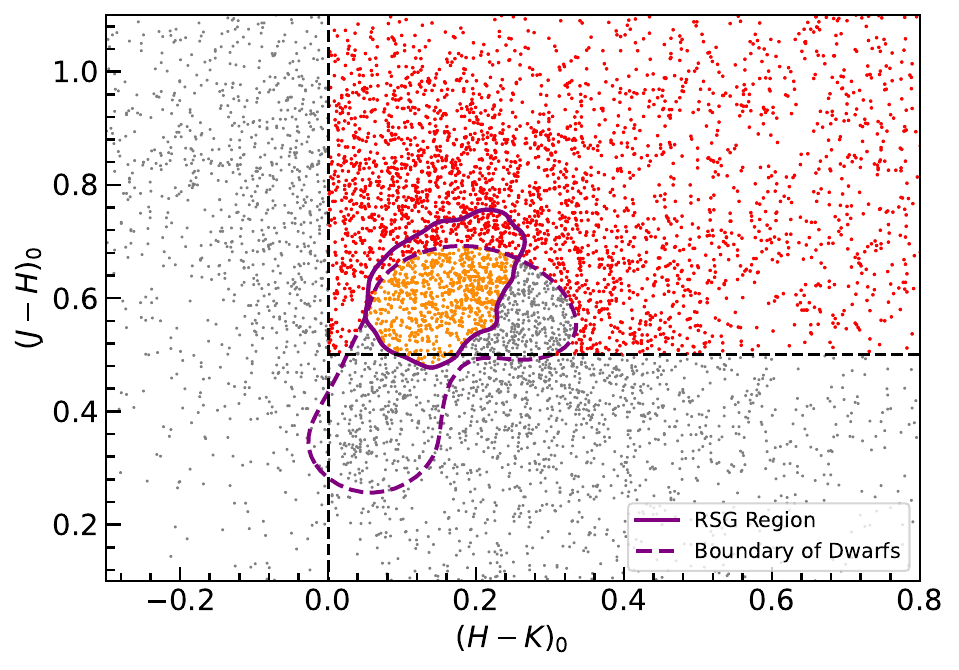}}
    \caption{The same as Figure \ref{Figure.rzh_selection}, but for the $(J-H)_0/(H-K)_0$ diagram. These are for NGC 6822, but with the SMC regions highlighted.}
    \label{Figure.jhk_selection}
\end{figure}

A pure but incomplete sample of RSGs is selected from the complete sample, which attempts to remove all the foreground dwarfs. During the selection of the complete sample, the empirical region of RSGs in the CCD is overlapped with that of the foreground dwarf stars (see Figure \ref{Figure.rzh_selection} and Figure \ref{Figure.jhk_selection}). Inevitably, some foreground dwarfs blend into our RSG candidates.
For the pure sample, the CCD criteria are modified to avoid dwarf stars blend into the RSG candidates. The empirical region of RSGs criterion is no longer used, instead all sources within the dwarf branch are removed. Only sources outside the dwarf branch are kept in the CCD. As shown in Figure \ref{Figure.rzh_selection} and Figure \ref{Figure.jhk_selection}, the red dots represent the pure sample, while the orange dots represent the sources that are removed compared to the complete sample.
Finally, the pure sample of member stars is consisted of 2,956, 5,288 and 2,720 sources that satisfy both the CCD and Gaia astrometric criteria, only the CCD criteria, and only the Gaia astrometric criteria, respectively.

\begin{figure}
    \centerline{\includegraphics[width=0.79\linewidth]{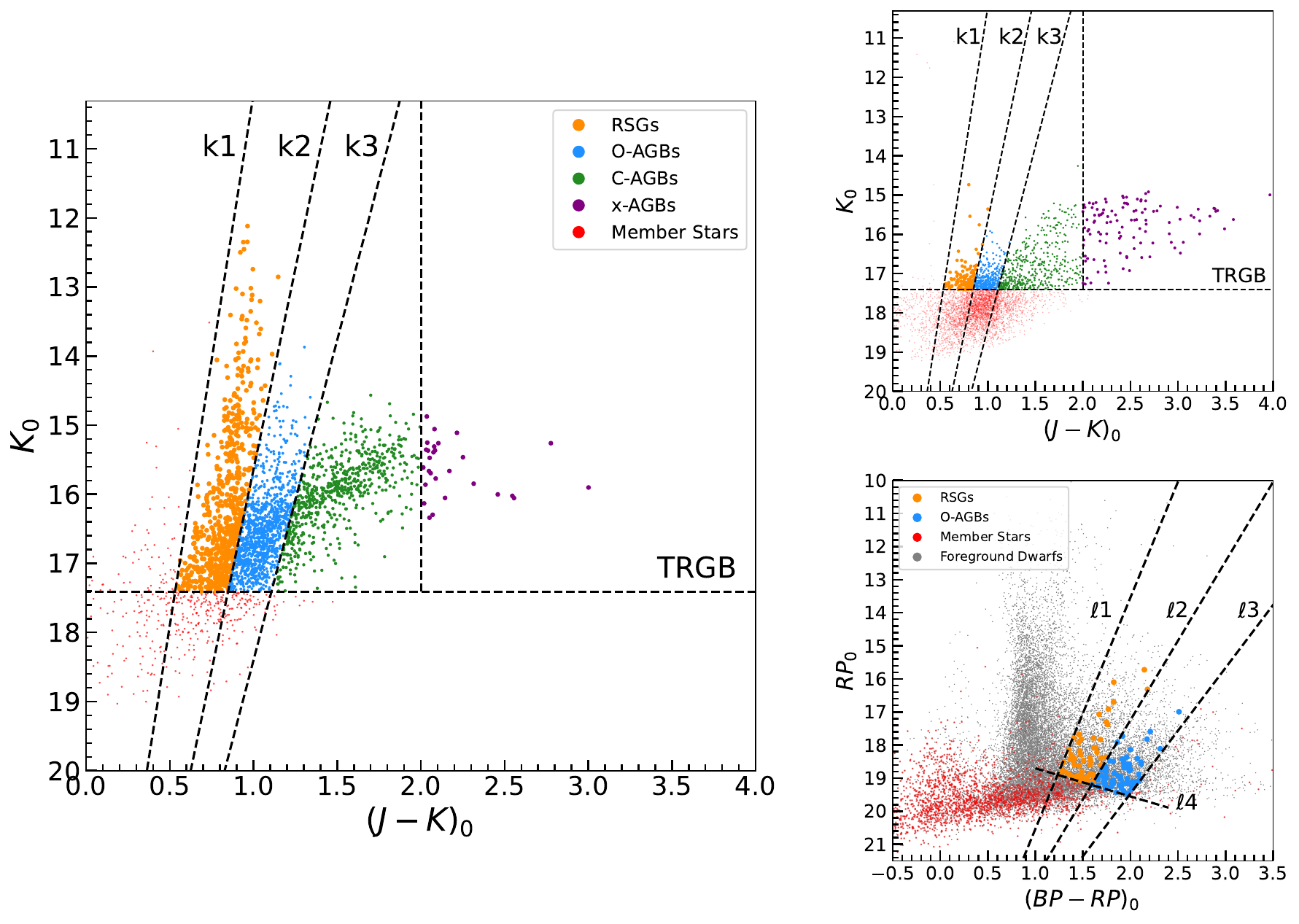}}
    \caption{Identified RSG and AGB candidates in the pure sample. The color convention is the same as in Figure \ref{Figure.complete_sample}.}
    \label{Figure.pure_sample}
\end{figure}

The pure sample of RSGs and AGBs are then identified in the CMD by the same criteria as for the complete sample. As a result, the pure sample has a total of 843 RSG candidates (660, 121, and 62 from the CCD-Gaia, CCD-only, and Gaia-only sample, respectively), 1,519 O-AGB candidates (1,123, 326, and 70 from the CCD-Gaia, CCD-only, and Gaia-only sample, respectively), 1,059 C-AGB candidates (612 and 447 from the CCD-Gaia and CCD-only sample, respectively) and 140 x-AGB candidates (28 and 112 from the CCD-Gaia and CCD-only sample, respectively). The distribution of RSG and AGB candidates of the pure sample in the CMD is shown in Figure \ref{Figure.pure_sample} and the number of identified RSGs and AGBs candidates are also listed in Table \ref{Table.final_sample}.

\section{Discussions}
\label{Sect.discussions}
In this section, we will analyze the contamination from foreground dwarfs and O-AGBs in the RSG sample, calculate the removal efficiency of the methods, and compare the results with previous works, and finally, estimate the number of newly identified RSGs in NGC 6822.

\subsection{Contamination Rate Estimation}
\subsubsection{Contamination From Foreground Dwarfs}
To estimate the contamination rate of RSGs and O-AGBs in the complete and pure sample by the foreground dwarfs, the same process is applied to the reference region where it's assumed that no RSGs or AGBs exist. The estimated contamination number of RSGs and AGBs in the complete sample and the pure sample is listed in Table \ref{Table.final_sample}.
Accordingly, the contamination rate, defined by the number of faked objects divided by the number of identified objects for a given type, is 20.5\%, 9.7\% , 6.8\%  and 5.0\% for RSG, O-AGB, C-AGB and x-AGB candidates in the complete sample respectively, and 6.5\%, 8.8\%, 6.1\% and 5.0\% respectively for the pure sample. It should be noted that the contamination rate from the foreground dwarfs is likely to be overestimated because the selected reference region is close to NGC 6822 and may include some member stars.

\subsubsection{Contamination From O-AGBs}
\label{Sect.Contamination From O-AGBs}
In the NIR bands, O-AGBs can be as bright as the faint RSGs. Though they are generally redder, the photometric uncertainty and extinction heterogeneity can bring them to the region of RSGs in the CMD, in particular those close to the borderline.
We examine the contamination of RSGs from O-AGBs by the NIR and optical CMD. 

\begin{figure}
    \centerline{\includegraphics[width=0.5\linewidth]{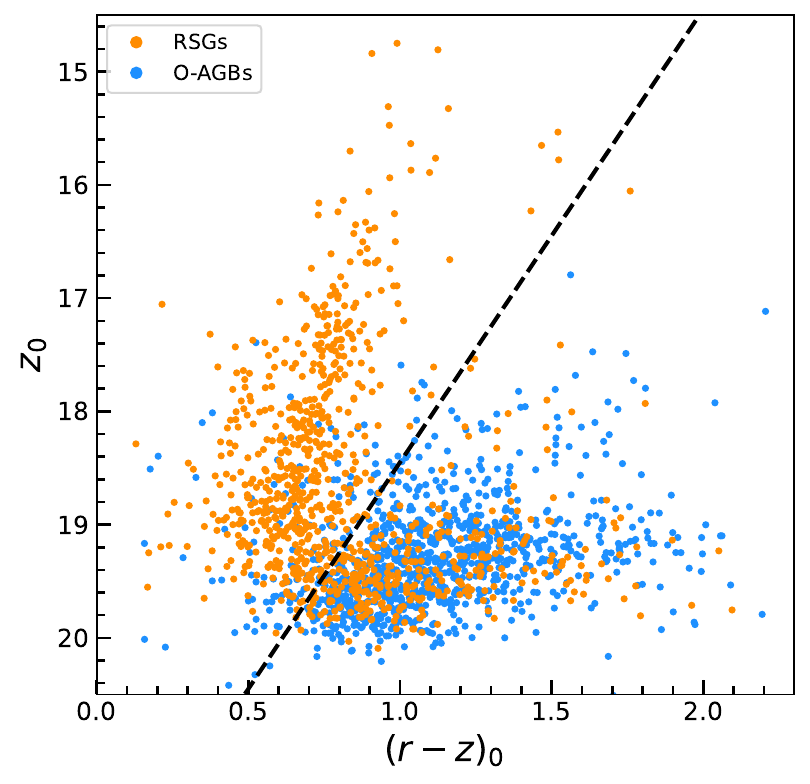}}
    \caption{The optical $(r-z)_0$ vs. $z_0$ diagram for RSGs (orange dots) and O-AGBs (blue dots) identified in the NIR CMD.
    }
    \label{Figure.optical_RSG}
\end{figure}

As shown in Figure \ref{Figure.optical_RSG}, the RSGs identified in the NIR CMD (orange dots) split into two branches in the optical $(r-z)_0$ vs. $z_0$ diagram. While the left branch coincides with the region of RSGs, the right branch belongs to the region of O-AGBs identified in the NIR CMD. We think the right branch stars are actually O-AGBs instead of RSGs.
The reason for such O-AGBs indistinguishable from RSGs in the NIR CMD is their relatively large photometric error that bring them into the RSG region. But in the optical bands, the circumstellar dust extinction of O-AGBs is much larger than the photometric uncertainty and thus leads to the apparent distinction in $(r-z)_0$.
In support, because the NIR photometry is much more precise due to the high brightness of the objects, the RSGs identified in the NIR CMD are not mixed with the O-AGBs in the optical CMD in the Magellanic Clouds (\citealt{2019A&A...629A..91Y, 2021A&A...646A.141Y, 2021ApJ...923..232R}).

In the complete sample of RSGs, there are a total of 1,184 RSG candidates identified in the NIR and Gaia CMD. Among them, 897 candidates have the optical data and are used to estimate the contamination rate from the O-AGBs.
As a result, 577 candidates are identified as RSGs and 320 as O-AGBs. Therefore, the contamination rate from the O-AGBs in the NIR CMD is about 35.7\%. In the pure sample, 665 out of 843 RSG candidates have optical data, of which 366 are identified as RSGs and 299 as O-AGBs, with a contamination rate from the O-AGBs in the NIR CMD of about 45.0\%. The increase of the contamination rate in the pure sample compared to the complete sample is due to fewer foreground dwarfs in the left branch being identified as RSGs.

\subsection{Efficiency of the Three Methods}
The relative efficiency in removing foreground contamination in this procedure is defined as the number of sources removed by one method diveded by total number of sources removed.
For the complete sample of member stars, the $(r-z)_{0}/(z-H)_{0}$ diagram, the $(J-H)_0/(H-K)_0$ diagram and the Gaia astrometric criteria remove 10,086, 3,625, and 5,281 sources among a total of 18,992 sources removed. The relative removal rates for the three methods are then 53.1\%, 19.1\%, and 27.8\%, respectively. For the pure sample of member stars, the corresponding numbers are 11,912, 4049, and 5281 sources, respectively, with a total of 21,242 sources removed. Therefore, the relative removal rates are 56.1\%, 19.1\%, and 24.9\%, respectively. Clearly, the $(r-z)_{0}/(z-H)_{0}$ diagram method is significantly more efficient than the other two methods in this procedure, as the foreground contamination in this work is firstly removed with this diagram.

\subsection{Comparison with Previous Works}
\begin{figure}
    \centerline{\includegraphics[width=1\linewidth]{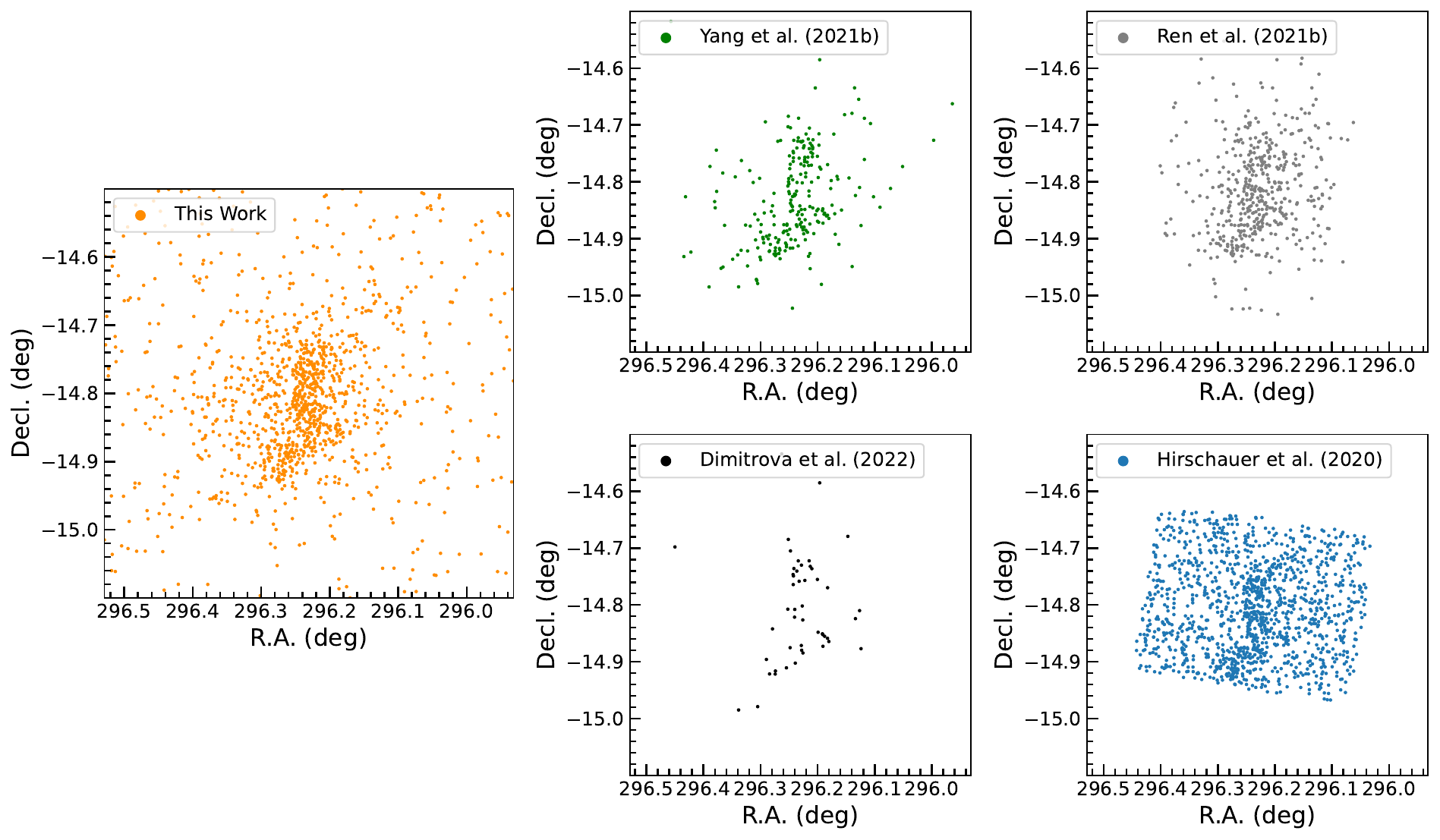}}
    \caption{The spatial distribution of the RSG candidates in this work, \citet{2021A&A...647A.167Y}, \citet{2021ApJ...923..232R}, \citet{2022AJ....163...70D}, and \citet{2020ApJ...892...91H}.}
    \label{Figure.RSGs_comparison}
\end{figure}

There are two relevant works on the identification of RSGs in NGC 6822 recently. One is \citet{2021A&A...647A.167Y} that identified 234 RSG candidates mainly by using the $(r-z)_{0}/(z-H)_{0}$ diagram, and the other is \citet{2021ApJ...923..232R} that identified 465 RSG candidates mainly by using the $(J-H)_0/(H-K)_0$ diagram. This work however identifies 1,184 and 843 RSG candidates in the complete and pure sample, respectively, both of which are significantly more numerous than either of the previous works. The spatial distributions of the RSG candidates in our work (orange dots), \citet{2021A&A...647A.167Y} (green dots), and \citet{2021ApJ...923..232R} (gray dots) are shown in Figure \ref{Figure.RSGs_comparison}.

The crossmatching of our catalog with \citet{2021A&A...647A.167Y} yields 213 and 179 common RSG candidates for the complete and pure sample, respectively.
For the complete sample, 21 RSG candidates identified by \citet{2021A&A...647A.167Y} are not in our sample. Among them, 14 sources are classified as other types of stars in our work due to the slightly different boundaries in the CMD, which means they are still in our sample of member stars. Seven sources are removed by our criteria, among which four and three are removed by the $(r-z)_{0}/(z-H)_{0}$ and $(J-H)_0/(H-K)_0$ diagram, respectively.
Furthermore, 34 RSG candidates are not in our pure sample because they are mixed into the dwarf branch and thus removed. 

Similarly, the crossmatching with \citet{2021ApJ...923..232R} yields 297 and 292 common RSG candidates for the complete and pure sample, respectively. For the complete sample, 168 sources identified by \citet{2021ApJ...923..232R} are not in our sample. Among them, 87 sources are classified as other types like AGBs or RGBs in the CMD due to either slightly different boundaries in the CMD, or different extinction correction. \citet{2021ApJ...923..232R} corrected the extinction independently for each source using the extinction map of \citet{2019ApJ...887...93G}, while our work correct the extinction uniformly by the central value in the extinction map of NGC 6822 by \cite{1998ApJ...500..525S}. In addition,
81 sources are removed as foreground dwarfs by our criteria, among which 34 and 3 sources are removed by the $(r-z)_{0}/(z-H)_{0}$ and $(J-H)_0/(H-K)_0$ diagram, respectively, and 44 by the Gaia astrometric criteria.
In addition, 5 RSG candidates are not in our pure sample because they are mixed into the dwarf branch.

Besides the above two works, other efforts have also been put to identify RSGs in NGC 6822. 
\begin{itemize}
\item \citet{2022AJ....163...70D} identified the brightest 51 RSG candidates in NGC 6822, by removing foreground contaminants with the Gaia/DR2 astrometric data. Their spatial distribution is shown by the black dots in Figure \ref{Figure.RSGs_comparison}. Among them, 46 are matched with our complete RSGs sample, and 5 are not, in which 1 is removed by the $(r-z)_{0}/(z-H)_{0}$ diagram, 1 by the Gaia astrometric criteria, and 3 lie outside our studied region. Furthermore, 2 RSG candidates are not in our pure sample because they are within the dwarf branch.
\item \citet{2020ApJ...892...91H} directly selected RSG candidates in NGC 6822 by using three CMDs. This sample contains 1,292 RSG candidates, and their spatial distribution is shown by the blue dots in Figure \ref{Figure.RSGs_comparison}, with a contamination rate of 52.93\% reported in the study. Among them, 350 and 270 RSG candidates are matched with the complete and pure sample, respectively. For the complete sample, 942 sources identified by \citet{2020ApJ...892...91H} are not matched, in which (1) 83 sources are classified as other types of stars, (2) 852 sources are removed by our criteria (532, 35, and 285 by the $(r-z)_{0}/(z-H)_{0}$, $(J-H)_0/(H-K)_0$, and Gaia astrometry criteria respectively), and (3) 7 sources are not included in our initial sample after the data quality control. In comparison with our pure sample, 80 RSG candidates in \citet{2020ApJ...892...91H} are removed because they are in the dwarf branch.
\item \citet{2022ApJ...933..197T} identified RSGs and AGBs candidates by using 3D color-color-magnitude diagram to remove foreground contaminants. Their sample contains a total of 2,788 C-AGB and O-AGB candidates, with a mean ratio between carbon and oxygen (C/M) stars of 0.67 $\pm$ 0.08. In our complete sample, there are a total of 2,634 C-AGB and O-AGB candidates, with a mean C/M ratio of about 0.69, consistent with theirs. This value is also consistent with $0.674^{+0.301}_{-0.145}$ from \citet{2022Univ....8..465R}. Further comparisons of the RSGs will be conducted once their data are publicly available.
\end{itemize}

In summary, the sample from \citet{2020ApJ...892...91H} has the largest number (350) of identical RSG candidates with our complete sample, and the sample from \citet{2021ApJ...923..232R} has the largest number (292) of identical RSG candidates with our pure sample. Although the sample from \citet{2020ApJ...892...91H} has slightly more common RSG candidates than from \citet{2021ApJ...923..232R}, their intrinsic contamination rate is apparently higher. 
Therefore, we provide an estimation of the newly identified RSG candidates in NGC 6822 in our work by comparing with \citet{2021ApJ...923..232R}, taking into account both the completeness and the purity. There are 297 and 292 identical RSG candidates between our work and \citet{2021ApJ...923..232R} in the complete and pure sample, respectively.
For the complete sample of 1,184 RSG candidates, considering a contamination rate of 20.5\%, approximately 600 new RSG candidates are identified. For the pure sample of 843 RSG candidates, considering a contamination rate of 6.5\%, approximately 450 new RSG candidates are identified. 
Compared to \citet{2021ApJ...923..232R}, our pure sample retains more RSG candidates, which are primarily selected by the $(r-z)_{0}/(z-H)_{0}$ diagram. In the $(r-z)_{0}/(z-H)_{0}$ diagram, RSGs are further from the dwarf branch than in the $(J-H)_0/(H-K)_0$ diagram, allowing more potential RSGs to be retained.
It should be noted that this estimation only considers contamination from foreground dwarfs, without accounting for contamination from O-AGBs. This is because that contamination from O-AGBs is intrinsic and inevitable due to photometric error and reddening, which are present in previous RSG samples as well. However, the complementary optical data can help mitigate this contamination as discussed in Sect.\ref{Sect.Contamination From O-AGBs}.

These comparisons indicate that the method used for identification of RSG candidates influences the sample significantly. The key factors include the data quality, foreground contamination removal, non-uniform extinction the definition of the RSGs and AGBs in the CMDs.

\section{Summary}
\label{Sect.conclusions}
The previous CCD method is metalicity-limited due to that the faint end RSGs will mix into the foreground dwarf branch as metallicity decreases, leading to the excessive removal. Meanwhile, Gaia astrometric method is distance-limited. This work attempts to create a new way to select a complete and pure sample of RSGs in nearby metal-poor galaxies. For this purpose, the region of RSGs in the $(r-z)_{0}/(z-H)_{0}$ and $(J-H)_0/(H-K)_0$ diagram is empirically defined and modified by fitting the locations of RSGs in the Magellanic Clouds, M31 and M33 with a wide range of metallicity from -1.0 to +0.3. Consequently, potential RSGs that are mixed into the dwarf branch are retained. Besides, we also combine the CCD method with the Gaia astrometry to remove the foreground contaminants. Among the three methods, the $(r-z)_{0}/(z-H)_{0}$ diagram method has the highest efficiency in this work.

NGC 6822 is taken as a case study by using this method. As a result, 1,184 RSG, 1,559 O-AGB, 1,075 C-AGB, and 140 x-AGB candidates are identified in the complete sample, with the contamination rate of approximately 20.5\%, 9.7\%, 6.8\%, and 5.0\%, respectively. Besides, a pure sample of RSGs and AGBs is selected from the complete sample by removing all the potential foreground sources. Consequently, 843 RSG, 1,519 O-AGB, 1,059 C-AGB, and 140 x-AGB candidates are retained in the pure sample, with the significantly reduced contamination rate of approximately 6.5\%, 8.8\%, 6.1\%, and 5.0\%, respectively.
By comparison with the RSGs sample of \citet{2021ApJ...923..232R}, about 600 and 450 RSG candidates are newly identified in our complete and pure sample, respectively.

In the future, we plan to apply this approach to more metal-poor and distant galaxies in the Local Group to identify the RSGs as complete as possible.

\section*{Acknowledgements}
We are grateful to Mr.\ Min Dai and Mr.\ Zehao Zhang for their friendly help and discussion. This work is supported by the National Natural Science Foundation of China through grant Nos.\ 12133002，12373048, and 12203025, National Key R\&D Program of China through grant No.\ 2019YFA0405503, China Manned Space Project through grant Nos.\ CMS-CSST-2021-A08 and CMS-CSST-2021-A09, Shandong Provincial Natural Science Foundation through project ZR2022QA064 and Shandong Provincial University Youth Innovation and Technology Support Program through grant No.\ 2022KJ138. This work has made use of the data from Gaia, PS1 and UKIRT.

\software{astropy \citep{2013A&A...558A..33A,2018AJ....156..123A},
          TOPCAT \citep{2005ASPC..347...29T}, dustmaps \citep{2018JOSS....3..695G}}

\bibliography{sample631}{}
\bibliographystyle{aasjournal}

\begin{table}[]
\centering
\caption{Number of RSGs and AGBs candidates in different samples and regions.}
\label{Table.final_sample}
\setlength{\tabcolsep}{5pt} 
\renewcommand{\arraystretch}{1.} 
\hspace*{-3cm} 
\begin{tabular}{@{}c c c c c c c@{}}
\toprule
Final sample & Type & {\parbox{2.5cm}{\centering CCD-Gaia\\sample }} & {\parbox{2.5cm}{\centering CCD-only\\sample }} & {\parbox{2.5cm}{\centering Gaia-only\\sample }} & Total & Contamination rate \\
\midrule
\multirow{4}{*}{Complete sample}  
  & RSGs    & 959   & 163  & 62  & 1,184 & 20.5\%  \\
  & O-AGBs  & 1,154 & 335  & 70  & 1,559 & 9.7\%   \\
  & C-AGBs  & 620   & 455  & -   & 1,075 & 6.8\%   \\
  & X-AGBs  & 28    & 112   & -   & 140   & 5.0\%   \\
\midrule
\multirow{4}{*}{\parbox{4cm}{\centering Reference region \\ }}  
  & RSGs    & 193   & 37   & 13  & 243   & -       \\
  & O-AGBs  & 56    & 84   & 11  & 151   & -       \\
  & C-AGBs  & 46    & 27   & -   & 73    & -       \\
  & X-AGBs  & 0     & 7    & -   & 7     & -       \\
\midrule
\multirow{4}{*}{Pure sample}  
  & RSGs    & 660   & 121  & 62  & 843   & 6.5\%   \\
  & O-AGBs  & 1,123 & 326  & 70  & 1,519 & 8.8\%   \\
  & C-AGBs  & 612   & 447  & -   & 1,059 & 6.1\%   \\
  & X-AGBs  & 28    & 112   & -   & 140   & 5.0\%   \\
\midrule
\multirow{4}{*}{\parbox{4cm}{\centering Reference region \\ }}  
  & RSGs    & 29    & 13   & 13  & 55    & -       \\
  & O-AGBs  & 45    & 78   & 11  & 134   & -       \\
  & C-AGBs  & 42    & 23   & -   & 65    & -       \\
  & X-AGBs  & 0     & 7    & -   & 7     & -       \\
\bottomrule
\end{tabular}
\end{table}




\end{CJK*}
\end{document}